# The Sunspot Catalogues of Carrington, Peters, and de la Rue: Quality Control and Machine-readable Versions

R. Casas[1] and J. M. Vaquero[2]

[1]Institut de Ciències de l'Espai (IEEC-CSIC), Campus UAB, Torre C5 parell 2n pis, E-08193 Bellaterra, Spain, email: casas@ice.cat

[2]Departamento de Física, Centro Universitario de Mérida, Universidad de Extremadura, Avda. Santa Teresa Jornet 38, E-06800 Mérida, Spain, email: jvaquero@unex.es

**Abstract**

In the 19th century, several astronomers made observations of sunspots, recording their positions and sometimes their areas. These observations were published in the form of extensive tables, but have been unhelpful until now. Three of these observers were Richard C. Carrington, Christian H. F. Peters, and Warren de la Rue (and their respective collaborators). They published, in various articles the data corresponding to 26 641 sunspot positions (Carrington, Peters, and de la Rue registered 4 900, 14 040, and 7701 sunspot positions, respectively). In this paper we present a translation of more than 400 pages of their printed numerical tables into a machine readable format, including an initial analysis targeted at detecting possible mistakes in the reading or in the original transcription. The observations carried out by these three astronomers have been made available at the Centre de Donées Astronomiques de Strasbourg (http://cdsarc.u-strasbg.fr/cgi-bin/VizieR?-source=VI/138).

Keywords: Solar Cycle, Observations; Sunspots, Statistics

## 1. Introduction

Since the Sun is a living star, with changes at different time scales, our knowledge of its behaviour requires not only new observations but also the analysis (or re-analysis) of historical data with current technology (Vaquero and Vázquez, 2009). Searches for solar data have now extended back to the most ancient records (Clark and Stephenson, 1978; Wittmann and Xu, 1987; Yau and Stephenson, 1988; Vaquero, Gallego, and García, 2002), but systematic and continuous solar observations really began with the invention of the telescope in the early 17th century.

Some of the first systematic drawings of the Sun were made by Thomas Harriot (Shirley, 1983), Galileo Galilei (Galilei, 1613), Christopher Scheiner (Scheiner, 1630), and Johannes Hevelius (Hevelius, 1647) from 1610 to 1645. They have been analysed by different researchers (Eddy, 1976; Eddy, Gilman, and Trotter, 1977; Herr, 1978; Sakurai, 1980; Herr, 1980; Abarbanell and Wöhl, 1981; Yallop et al., 1982; Casas, Vaquero, and Vázquez, 2006) with the aim of improving our knowledge of solar rotation prior to an important event, the Maunder Minimum (1645-1715) (Eddy, 1977). Other isolated observations from this period have been analysed (Vaquero, Sánchez-Bajo, and Gallego, 2002; Casas, Vaquero, and Vázquez, 2006; Vaquero et al., 2011). Hoyt and Schatten (1998) reported to have found that some astronomers observed the Sun during the Maunder Minimum and confirmed the lack of spots in the Sun's photosphere during this period.

During the 19th century, systematic observations were made by specific astronomers and by observatories that provided drawings and data of sunspot positions, with evaluations of their areas, and included the first photographs of the Sun (Rothermel, 1993; Bonifácio, Malaquias, and Fernandes, 2007). Between 1825 and 1867, the German astronomer Samuel Heinrich Schwabe made 8486 drawings of the full solar disc populated with spots. These data, preserved in the archives of the Royal Astronomical Society (London), have been digitized for future analyses (Arlt, 2011; Diercke, Arlt, and Denker, 2012).

The activity index, introduced by Rudolf Wolf in 1848, was reconstructed by him back to the first telescopic observations. His data are now available on various web sites (Clette et al., 2007; Vaquero, 2007). Other astronomers who observed the Sun in the 19th century and determined sunspot positions were Richard C. Carrington, Christian H. F. Peters, and Warren de la Rue (and their many collaborators). The recovery of the observations published by these three astronomers is the main objective of the present work.

In Section 2, we shall briefly describe the data published by these three astronomers and the methods they used. Section 3 describes the digitization of the data and the quality control applied in the process, explaining how we compared the data to reveal possible mistakes in the digitization or in the original transcription of the data. Finally, Section 4 summarizes the results.

**2. Observations**

Richard C. Carrington, assisted by various collaborators whom he cited in his paper (Carrington, 1863; Cliver and Keer, 2012), observed the Sun from the Redhill Observatory, South London (UK), from 1853 to 1861. The observatory was equipped with an equatorial telescope of 4.5 inches aperture and 52 inches focal length. To observe the Sun, he placed two perpendicular gold wires in the focal plane, at 4 from the East-West direction. The image of the Sun and the reticle was projected by a 25-power eyepiece onto a screen, obtaining an image of 12-14 inches at the beginning which was reduced to 11 inches later. The method of determining the sunspot positions is described in two of his papers (Carrington, 1853; Carrington, 1863) and has been reviewed relatively recently (Teague, 1996). In nine years, Carrington determined 4900 positions of sunspots, publishing this data in a table displayed over 101 pages in Section II of his book (Carrington, 1863). The data for each entry are: date, day of the year in decimal format, sequential observation number, polar position of the sunspot represented with its distance to the solar disc centre (using the solar radius as the unit) and position angle from celestial North to East, first node, heliographic longitude and latitude and group number.

Christian H. F. Peters made his observations between 1860 and 1870 from the Hamilton Observatory, in Clinton, New York (USA) (Bruce, 1987). He used a 13.5-inch Spencer refractor at f/13.3. His observations were published with the title "Heliographic positions of sun-spots observed at Hamilton College from 1860 to 1870" in 1907 by Edwin B. Frost after Peters died (Peters, 1907). Frost used Peters' notes and papers describing the observation and reduction methods. Peters' legacy comprises 14 040 sunspot positions covering a table of 170 pages. For each day, a title indicates the date and the astronomical mean time for the civil date. A letter in the first column indicates the spot and a letter in the last column indicates the spot on the next day. The data for each sunspot consists of the right ascension and declination differences with respect to the centre of the solar disc, the heliographic longitude and latitude, and the longitude of the node.

Warren de la Rue observed from the Kew Observatory (London, UK) during the period 1862-1866. He used a photo-heliograph designed by Sir John Herschel with an aperture of 3.4 inches and a focal length of 50 inches (Chree, 1916). De la Rue used the pictures obtained with the photo-heliograph to determine the positions and the areas of the sunspots, and published them in two papers (de la Rue, Steward, and Loewy, 1869; de la Rue, Steward, and Loewy, 1870). A total of 7701 sunspots were registered during his campaign.

All four papers were found to be available in digital format. Carrington's paper (Carrington, 1863) can be found on Internet Archive[1]. Peters' observations (Peters, 1907) are to be found on the same web site[2], and the two papers of de la Rue (de la Rue, Steward, and Loewy, 1869; de la Rue, Steward, and Loewy, 1870) are available on the web site of the Philosophical Transactions of the Royal Society of London[3,4].

We used two different methods of digitizing numerical data from printed text for the de la Rue data: optical character recognition (OCR) and key entry. Finally, all the tables were read with commercial OCR software according to the characteristics of the available files (Brönnimann et al., 2006). Basically, two problems were found when reading the tables: (i) the quality of the original paper and the scanned document, and (ii) the typography used in the original papers.

**3. Data Verification and Quality Control**

As mentioned above, the read-out of the tables with OCR software can introduce various errors. In addition, it is also possible that the original papers contain wrong values. We therefore attempted to find both types of error by crosschecking the internal consistency of the data given by the authors. In their tables, Carrington and de la Rue included the heliographic longitude, the heliographic latitude, the distance of the structure to the disc centre (in units of solar radius), and the position angle with respect to the celestial North. In addition to the heliographic longitude and the latitude, Peters included the distance of the sunspot to the solar centre for both axes (hourly angle and declination), the former in seconds and the latter in arc seconds.

---

[1] http://archive.org/details/observationsofsp00carr
[2] http://archive.org/details/heliographicposi00peterich
[3] http://rstl.royalsocietypublishing.org/content/159/1.full.pdf
[4] http://rstl.royalsocietypublishing.org/content/160/389.full.pdf

The complementary data allows one to compute the heliographic latitude and longitude of each spot and compare it with the values given by the authors. To do this, we needed the physical ephemerides for the Sun for each observation. These ephemerides are: latitude ($B_0$) and longitude ($L_0$) of the subsolar point, the apparent solar radius (or diameter), and the position angle of the rotation axis ($P_0$).

3.1. Ephemerides sources and comparison

To obtain these ephemerides, we considered two different methods. Firstly, we used the Interactive Data Language (IDL) code sun.pro written by R. Sterner in 1991, based on Meeus's book (Meeus, 1983). Secondly, we took the ephemerides from the Horizons web page (http://ssd.jpl.nasa.gov/horizons.cgi) (Giorgini et al., 1996).

We then compared the two sets of ephemerides for a period of time corresponding to the epoch of the observations. Figure 1 shows the differences between the two sets, i.e., the Horizons values minus the IDL code values. The central meridian presents a constant difference of (0.13±0.01)º, while the latitude shows an annual variation with an amplitude of 0.04º.

Despite the IDL code being able to provide values for any given time, it might have less precision than Horizons. For this reason, we here use this latter source to compute the Sun's physical ephemerides. With the Horizons web interface, we obtained a table with all the data needed at 00:00 UT, interpolating to the time of each observation.

3.2. Carrington's observations

As Carrington's observations were made near London, we considered the times included in his observations to correspond already to UT, so that no corrections were needed. We evaluated the heliographic longitude and latitude for each observation from the distance to the solar disc centre and the position angle, including the physical ephemerides obtained by interpolating the Horizons ephemerides to the observing time. We then subtracted the two sets of data to compare the calculated heliographic coordinates with the values obtained by Carrington. From the mean difference and standard deviation (σ) of the longitude and the latitude, we checked all the data that exceeded 3σ, and corrected the value either if it was an OCR software mistake or if there was a very clear typographical mistake in the original paper.

Having applied these corrections, we then re-calculated the differences between Carrington's values and our values. These values are summarized in Table 1 and Figure 2. The difference of -7.99º is large, but we do not know its origin. The three peaks in the latitude distribution could be due to data rounding by Carrington.

3.3. Peters' observations

The comparison to Peters' observations turned out to be more complicated. There were several reasons for this:

- The time published by Peters is (sic) "Clinton mean time (astronomical) for the civil date". Our interpretation of this sentence, explained by a histogram of the observing times, is that the time is referenced to noon, so that one should add 12 hours to the tabulated time, and that one should consider the geographical longitude of Clinton city ($5^h\ 1^m\ 37^s 5$ W) as given in Peters (1907, page iii).
- The heliographic longitudes are counted in the opposite direction to the rotation of the Sun (Peters, 1907, p. vii) and in the reverse direction to the angles measured by Carrington. Fortunately, the paper included a table calculated by Mr. Philip Fox with the angular distance between the prime meridians of Peters and Carrington fro Clinton noon for each day on which Peters carried out an observation. We made a straight line fit to these data to convert Carrington's longitude, which we calculated, to Peters' longitude. The fit is shown by Equation (1):

$$L_{0,Carrington} + L_{0,Peters} = (361625.85 \pm 0.20) - (0.150403810 \pm 10^{-8}) \cdot JD, \quad (1)$$

  where $L_{0,observer}$ is the prime meridian longitude for each observer in degrees and JD is the Julian date.

- The Cartesian coordinates of the sunspots calculated by Peters are in arc seconds in the declination and in seconds in the hourly angle. Therefore, one needs to know the declination of the solar disc centre. We used the Horizons web interface to calculate this value.

The analysis of the Peters' data gave similar mean differences as that of Carrington's data, but now with larger standard deviations (see Table 2). The different sign in the longitude difference is due to the opposite direction of Peters' longitude. The histograms of the data (Figure 3) are more clearly single-peaked, although with large standard deviations.

3.4. de la Rue's observations

Since the format of de la Rue's data is the same as that of Carrington's observations, and they were obtained at fixed observatories, we used the same procedure to compare his data with ours. De la Rue's data present various peculiarities. Figures 4 and 5 show the differences between de la Rue's heliographic longitudes and latitudes and our calculated values. Despite the presence of outliers, there is a clear increase in the longitude dispersions from 1 January 1864 (JD 2401871.5) onwards. However, the most striking feature of these figures is that, from this day onwards, the latitude presents a sinusoidal behaviour with a period of 1 year and an amplitude of 14.5º.

We subtracted our calculated values of B0 from each of de la Rue's observations from 1 January 1864 onwards. The overall results are shown in Figure 6. There is still a major dispersion, although the sinusoidal behaviour has disappeared.

It is clear that the statistical analysis must consider two samples, i.e. that before 1 January 1864 and that from that date onwards. The corresponding median and median absolute deviation (MAD) values are listed in Table 3.

Figure 7 shows the histograms for the differences between the longitude and the latitude calculated by de la Rue and by us. It also includes the latitude of de la Rue corrected as has been explained previously in the text.

Finally, the scatter plots for each sample (Figure 8) show the spatial distribution for the differences.

**4. Conclusions**

Sunspot series and other one-dimensional index time-series have been widely used in solar physics studies. However, recent research, such as the development of advanced solar-dynamo models, requires information about the location and dynamics of active regions. The various sunspot catalogues available (Lefevre and Clette, 2012) are mainly restricted to the second half of the 20th century.

The main objective of this work has been to provide machine-readable versions of the sunspot positions determined by Carrington, Peters, and de la Rue in the second half of 19th century. We include a new file (combdata.dat) with the heliographic positions of sunspots calculated using the polar or cartesian coordinates given by Carrington, Peters and de la Rue in their papers and the ephemerides obtained from Horizons web page. This file provides uniform heliographic coordinates for the period covered by these observations. These data are now available at the Centre de Donées Astronomiques de Strasbourg[5] and on a personal web page[6], so that the solar physics community can incorporate historical information on solar activity into studies of the position of sunspots, including differential rotation, active longitudes, North-South asymmetries, etc. The data contained in these files are the original obtained by Carrington, Peters and de la Rue except for a few clear corrections listed in Table 4.

As an example of the intrinsic interest of these data, Figure 9 shows the butterfly diagram for the dates of the data retrieved in the present study. This diagram shows a good coverage of the end of Solar Cycle 9 (with sunspots close to the solar equator), the complete Solar Cycle 10, and the start of Solar Cycle 11 (with sunspots at high latitudes).

Our quality analysis checking the correctness of the digitization has shown that Carrington's data were of high quality. Peters' data, however, contained some clear outliers, and de la Rue's data presented an annual behaviour after 31 December 1863, reflecting a possible error in the application of the subsolar latitude $B_0$ in the calculation of the heliographic latitude.

---

[5] http://cdsarc.u-strasbg.fr/cgi-bin/VizieR?-source=VI/138
[6] http://segre.ice.cat/casas/sun/

**Acknowledgements**

Ricard Casas acknowledges the financial support received from the Spanish Ministerio de Ciencia e Innovación (MICINN), project AYA2009-13936, Consolider-Ingenio CSD2007-00060, and research project 2009SGR1398 from the Generalitat de Catalunya. José M. Vaquero acknowledges financial support from the Junta de Extremadura (Research Group Grant No. GR10131) and the Ministerio de Economía y Competitividad of the Spanish Government (AYA2011-25945).

Table 1. Mean values and standard deviations of the differences in the heliographic longitude (ΔL) and latitude (ΔB) between the coordinates evaluated by Carrington and by us.

|        | Mean ° | σ ° |
|--------|--------|-----|
| ΔL =   | −7.99  | ±0.14 |
| ΔB =   | −0.01  | ±0.14 |

Table 2. Differences between the heliographic coordinates calculated using the Cartesian coordinates of the sunspots and the values given by Peters.

|        | Mean ° | σ ° |
|--------|--------|-----|
| ΔL =   | 7.27   | ±3.01 |
| ΔB =   | 0.01   | ±1.57 |

Table 3. Median and MAD of the differences between the de la Rue's data and the heliographic coordinates evaluated by us after correcting his values by subtracting $B_0$ from the observations carried out after 31 December 1863.

|             | Longitude | | Latitude | |
|-------------|-----------|-----|----------|-----|
|             | Median ° | MAD ° | Median ° | MAD ° |
| Before 1864 | −7.50    | 0.03  | −0.01    | 0.01  |
| After 1863  | −7.70    | 3.09  | −0.06    | 2.01  |

Table 4. Modifications made in available files and original incorrect values.

| Obs. N. | Date | Title | Incorrect value | Modification |
|---------|------|-------|-----------------|--------------|
| **Carrington's data** | | | | |
| 480   | 30th March 1855    | No.              | 1480    | 0480    |
| 1183  | 13th July 1857     | H. Lat. (deg)    | −22     | −32     |
| 1185  | 14th July 1857     | H. Lat. (deg)    | −22     | −32     |
| 2988  | 10th November 1859 | H. Long. (deg)   | 279     | 269     |
| 2989  | 10th November 1859 | H. Long. (deg)   | 279     | 269     |
| 5276  | 23rd March 1861    | Date             | 23      |         |
| 5277  | 23rd March 1861    | Date             |         | 23      |
| **Peters' data** | | | | |
| p. 39 | 7th July 1861      | H. Long (deg min)| 32 429  | 324 29  |
| **de la Rue's data** | | | | |
| 594   | 5th May 1862       | H. Long. (deg)   | 535     | 335     |
| 1667  | 16th October 1862  | Mean Time        | 284.440 | 288.440 |
| 3525  | 22nd November 1863 | H. Long. (deg)   | 629     | 329     |
| 3592  | 12th December 1863 | H. Lat. (deg)    | −50     | −10     |

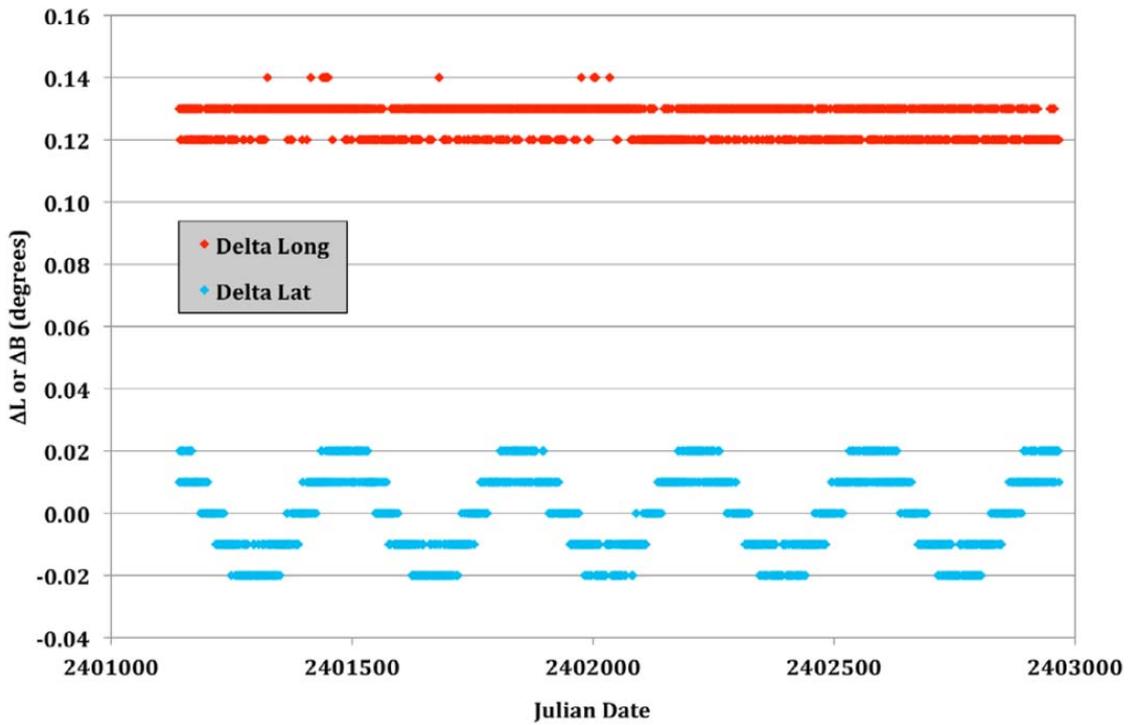

**Figure 1** Difference between the physical ephemerides of the Sun in the period of the observations presented in this study, using two methods (*Horizons* minus sun.pro).

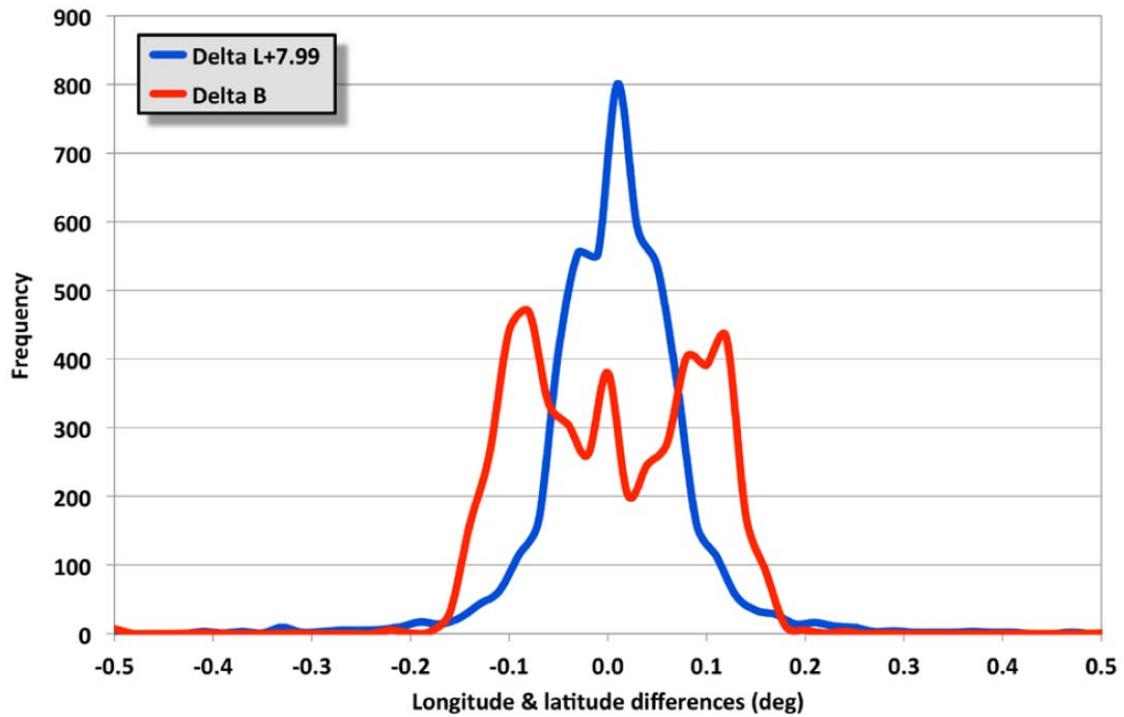

**Figure 2** Histograms for the differences between the heliographic longitudes and latitudes calculated by Carrington and by us. The longitude difference shows a sharp peak shifted by 7.99°, while the latitude shows a broader distribution with three clear peaks centred around the tenths of a degree.

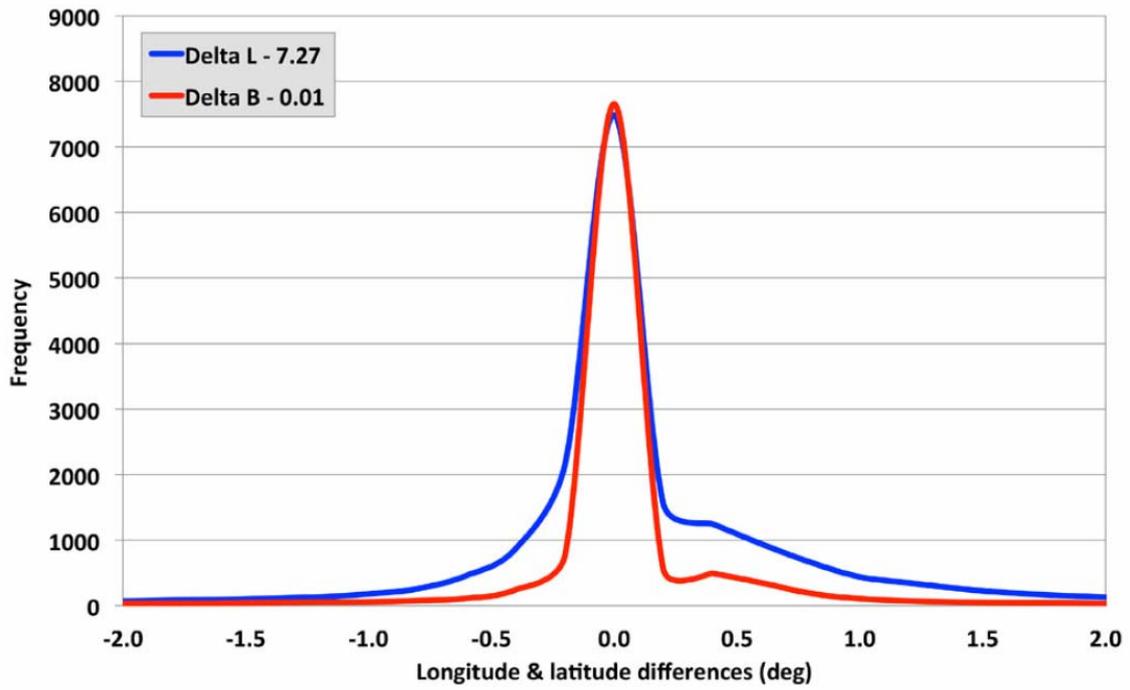

**Figure 3** Histograms of the differences between the longitude and latitude calculated by Peters and by us.

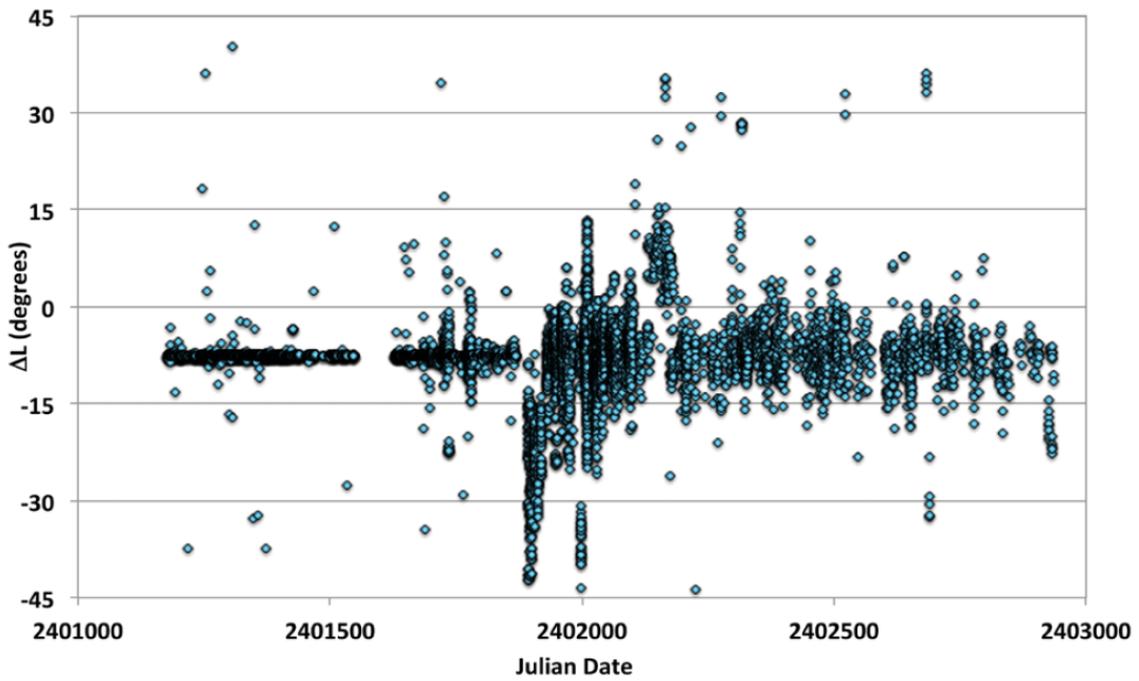

**Figure 4** Difference between the longitude calculated by de la Rue and by us. The dispersion increases considerably from 1 January 1864 (JD 2401871.5) onwards.

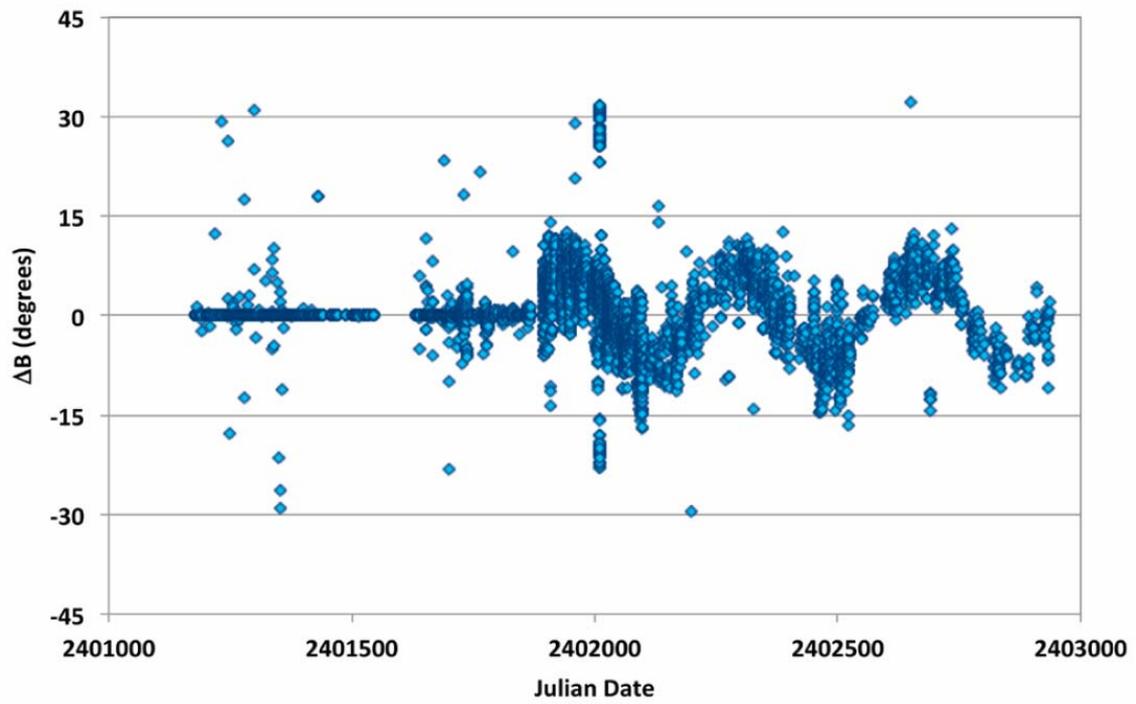

**Figure 5** Difference between the latitude calculated by de la Rue and by us. The behaviour is sinusoidal from 1 January 1864 with a period of a year and an amplitude of 14.5°.

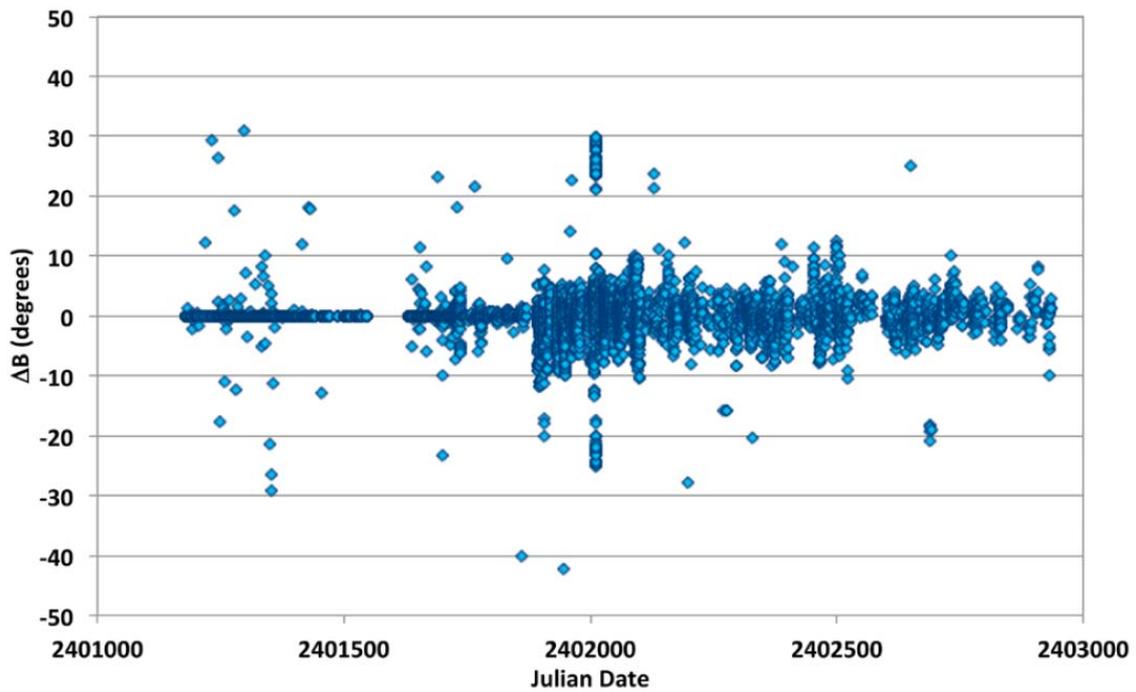

**Figure 6** Same as Figure 5 but removing the sinusoidal behaviour after 1 January 1864 by subtracting $B_0$.

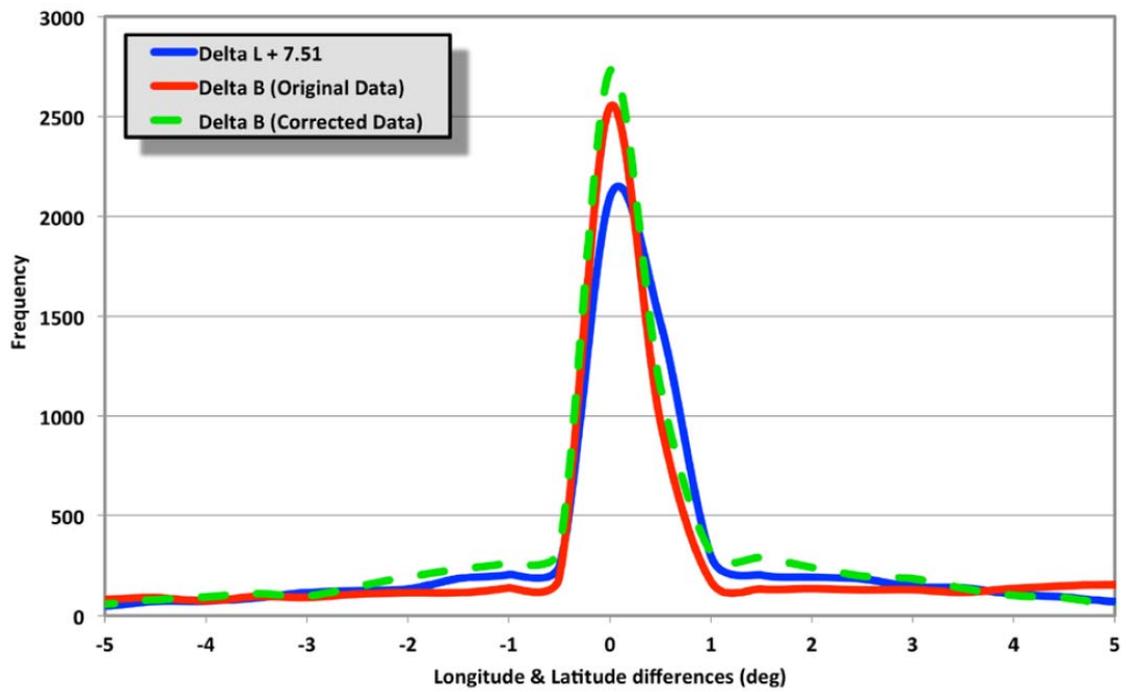

**Figure 7** Histograms of the differences between the longitude and latitude calculated by de la Rue and by us subtracting the median of the entire sample ($-7.51°$), with and without the correction of the latitude.

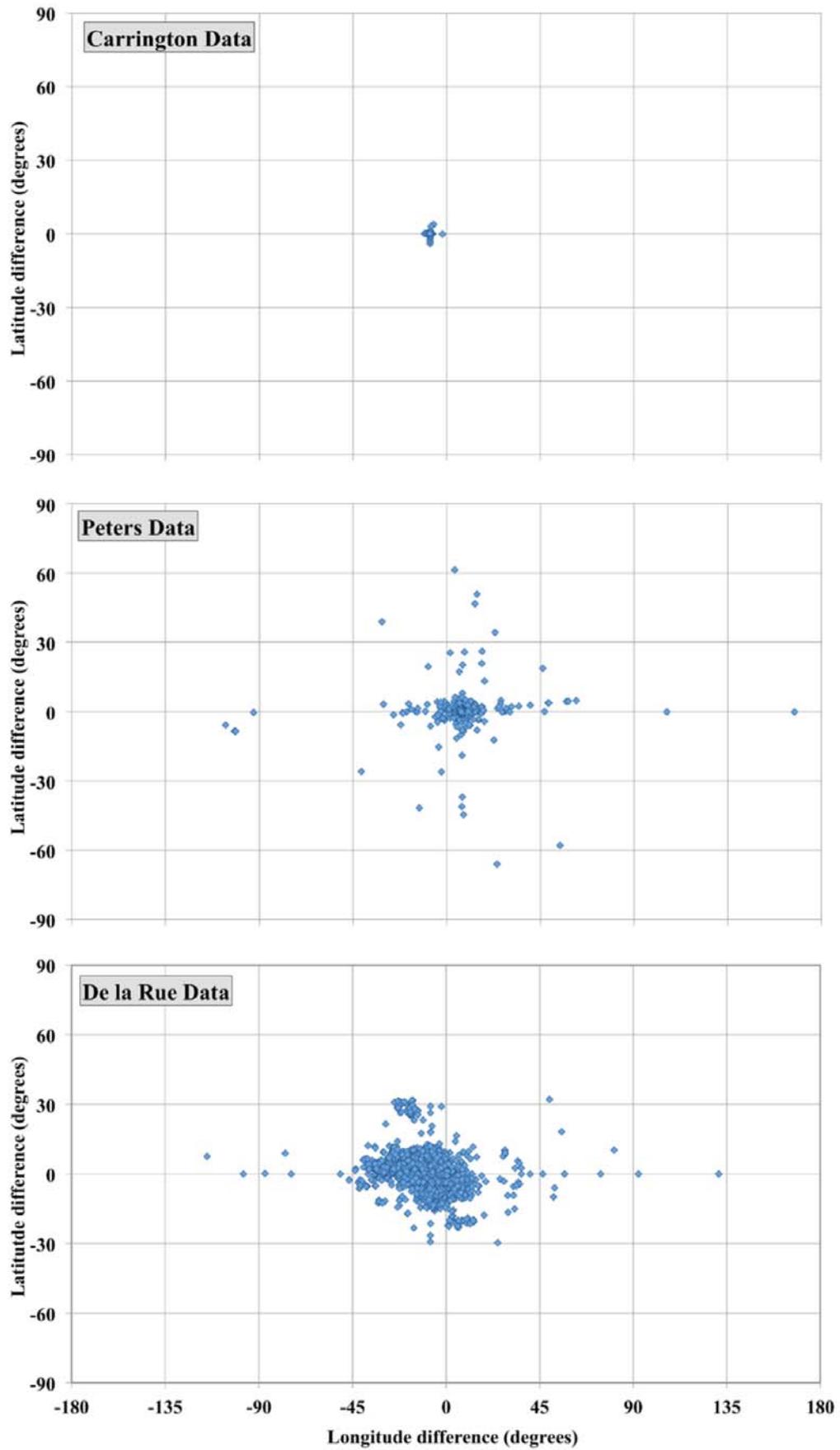

**Figure 8** Scatter plots for each set of observations with respect to the values calculated by us.